\documentclass[article]{elsarticle}

\usepackage{lineno,hyperref}
\modulolinenumbers[5]

\begin{document}

\begin{frontmatter}

\title{ALICE Measurements in p-Pb Collisions: Charged
Particle Multiplicity, Centrality Determination and
implications for Binary Scaling}

\author[mymainaddress,mysecondaryaddress]{Alberica Toia}
\ead{alberica.toia@cern.ch}
\address{on behalf of the ALICE Collaboration}
\address[mymainaddress]{Istituto Nazionale di Fisica Nucleare, Padova}
\address[mysecondaryaddress]{Goethe University Frankfurt}

\begin{abstract}
Measurements of particle production in proton-nucleus collisions
provide a reference to disentangle final state effects,
i.e. signatures of the formation of a deconfined hot medium, from
initial state effects, already present in cold nuclear matter. Since
many initial state effects are expected to vary as a function of the
number of collisions suffered by the incoming proton, it is crucial to
estimate the centrality of the collision.  In p-Pb collisions
categorization of events into different centrality classes using a
particle multiplicity distribution is complicated by the low particle
multiplicities and the large multiplicity fluctuations.  We present
ALICE measurements of particle production in p-Pb collisions at
$\sqrt{s_{NN}} = 5.02$ TeV, including the pseudo-rapidity and
transverse momentum dependence, we discuss the event classification in
centrality classes and its implications for the measurements of
nuclear modification factors.
\end{abstract}

\end{frontmatter}

\linenumbers

\section{Introduction}
Proton-lead collisions are an important component of the LHC heavy-ion
program for their reference role in the understanding and
interpretation of the nucleus-nucleus data, to disentangle final state
effects, signature of the formation of a hot QCD matter, from initial
state effects, already present in cold nuclear matter
\cite{Salgado}. 

Since many initial state effects are expected to vary as a function of
the impact parameter of the collision, it is crucial to estimate the
centrality-dependence of various observables, including multiplicity
and transverse momentum, and to categorize each event according to its
centrality. One then needs to determine $\ensuremath{N_\mathrm{coll}}$
for each centrality class.

In this proceeding we present and discuss the ALICE strategy to measure
centrality in p-Pb collisions and understand the dynamical bias
generated when ordering events according to their centrality.

ALICE has measured the nuclear modification factors $R_{pA}$ in minimum
bias collisions \cite{AliceRpA}, where the average p-Pb overlap
function $\langle T_{pA} \rangle$ is determined by total geometric p-A
cross-section. Nuclear effects in p-A collisions should be quantified
by a comparison of the p-A results to an incoherent superposition of
p-nucleon collisions. To make these measurement centrality-dependent,
event classes have to be defined using centrality estimators, that can
be either particle multiplicity or energy deposited in a given
pseudo-rapidity region. For each centrality class, two independent
questions need to be answered, namely: how many collisions
($\langle \ensuremath{N_\mathrm{coll}} \rangle$) occur in that sample? and how
unbiased are the nucleon-nucleon collisions?

To determine centrality we compare the signals in different ALICE
detectors, covering various rapidity regions, which are sensitive to
different kind of physics.  Particle production measured by detectors
around mid-rapidity can be modeled with a negative binomial
distribution, while the zero-degree energy of the slow nucleons
emitted in the nucleon fragmentation requires more sophisticated
models \cite{Sikler, Oppedisano}. The main estimators used for
centrality are:
\begin{itemize}
\item CL1 ($|\eta| < 1.4$): denotes the clusters measured in the Silicon Pixel detector
\item V0A ($2.8<\eta <5.1$): is the amplitude measured by the VZERO hodoscopes on the A-side (the Pb-remnant side),
\item V0M: is the sum of V0A ($2.8<\eta <5.1$) + V0C ($-3.7<\eta <-1.7$) ,
\item ZNA: is the energy deposited in Zero-degree neutron calorimeter on the A-side.
\end{itemize}
Using these estimators we are sensitive to the reaction products of
p-N collisions, the Pb fragmentation products that go mainly in the
direction of the Pb beam (towards V0A) and the so called slow nucleons from
evaporation and knock-out that are emitted into the very forward
directions and are detected by the zero degree calorimeters.

\section{Determination of $\ensuremath{N_\mathrm{coll}}$}

\subsection{NBD-Glauber fit for charged particle multiplicity}
\begin{figure}[btp]
 \centering
  \includegraphics[width=0.61\textwidth]{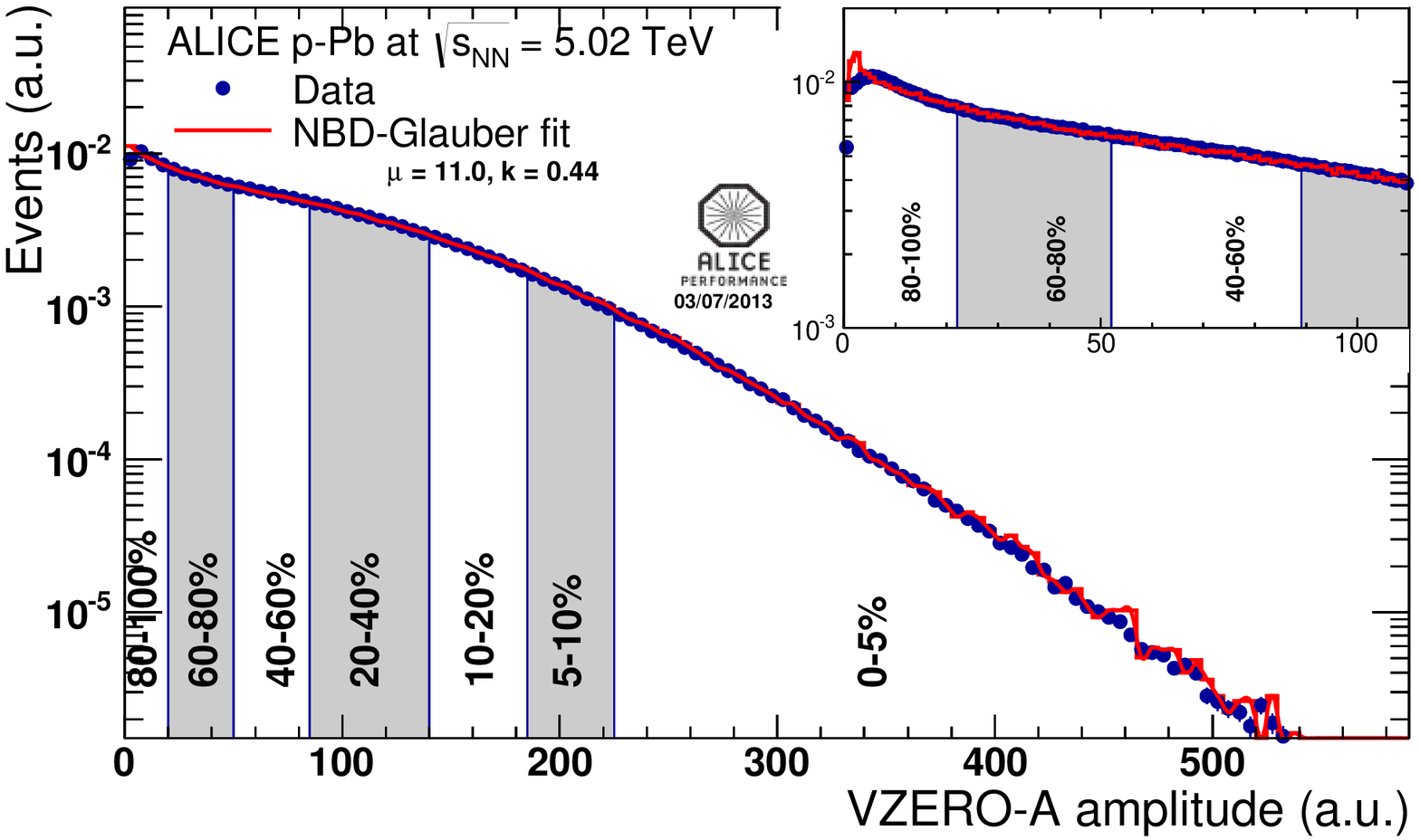}
  \includegraphics[width=0.38\textwidth]{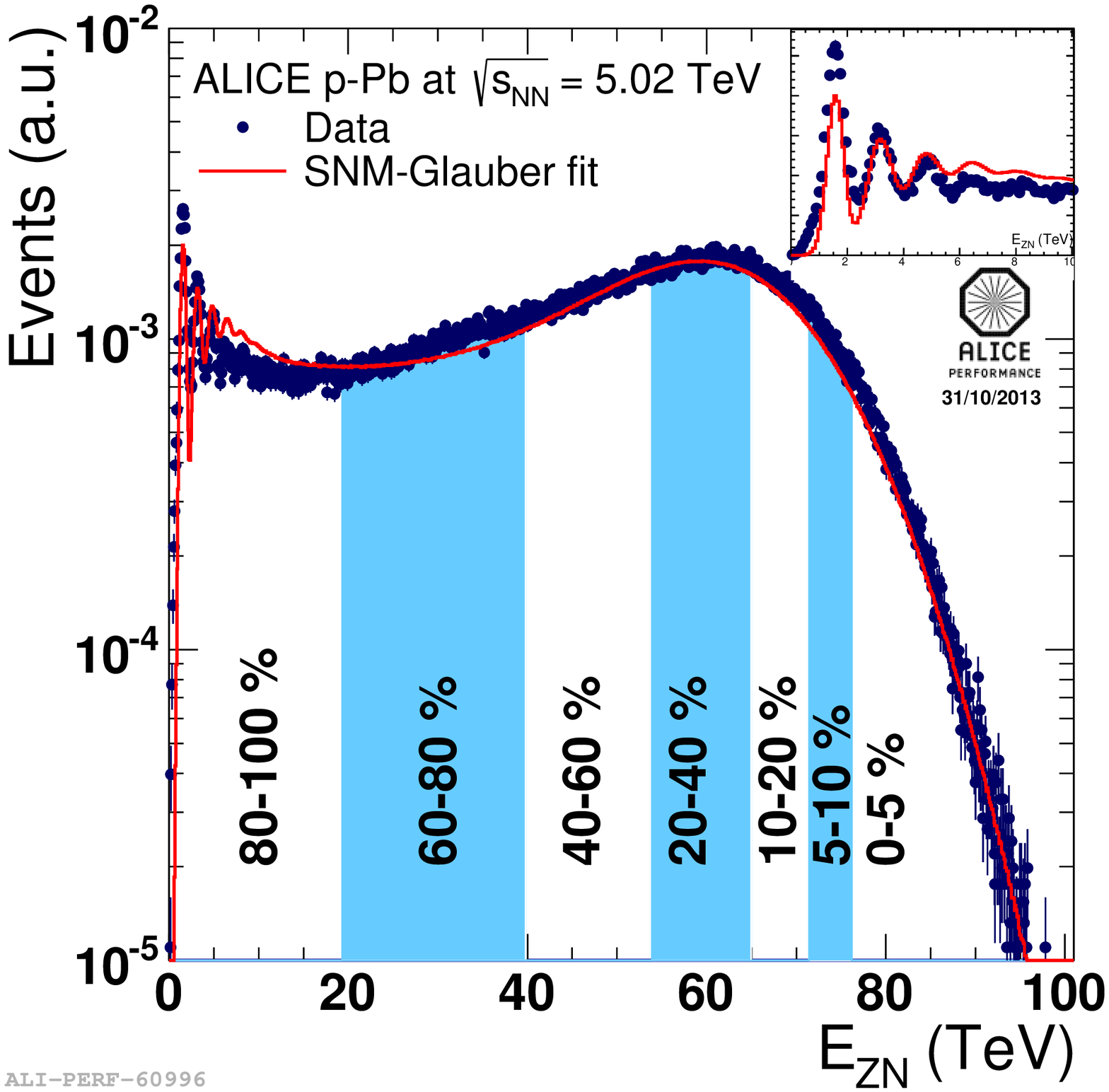}
  \caption{Left: Sum of amplitudes in the VZERO-A scintillators. Right: neutron energy spectra measured in the ZN-A calorimeter. The distributions are compared with the ones obtained from the NBD-Glauber and the SNM-Glauber fit respectively. Some centrality classes are indicated in the figure.
  \label{fig:glauFit}}
\end{figure}

For multiplicity measured by detectors around mid-rapidity, the
centrality is defined by an analogous procedure to what has been done
for Pb-Pb \cite{AliceCent}.  The measured multiplicity distribution is
divided in percentiles of the hadronic cross-section.  The
distribution P($\ensuremath{N_\mathrm{part}}$) is calculated with a
p-Pb Glauber-MC \cite{PHOBOS}. For each
$\ensuremath{N_\mathrm{part}}$, the multiplicity is calculated
according to a Negative Binomial Distribution (NBD). The NBD
parameters are fitted to the measured distribution. Then the
$<\ensuremath{N_\mathrm{coll}}>$ are calculated for each centrality
class. Figure \ref{fig:glauFit}, on the left, shows an example for
V0A. The same fit procedure is applied for the distribution of CL1 and
V0M. The obtained values for $\ensuremath{N_\mathrm{coll}}$ are
similar for different estimators. The systematic error was estimated
by varying Glauber MC parameters. We performed a MC closure test with
HIJING to confirm the correctness of the approach. The maximum
difference between the various estimators is smaller or consistent
with the established uncertainty and with the difference between using
a multiplicity estimators, or for centrality classes obtained by
dividing the impact parameter distribution in percentiles.

\subsection{SNM-Glauber fit for zero-degree energy}
With a totally different approach, the zero-degree energy measured in
the ZDC can also be used to extract the number of collisions. The ZDC
measures the slow nucleons emitted in the Pb-fragmentation
process. For this purpose, a more sophisticated model for particle
production, the Slow Nucleon Model (SNM) \cite{Sikler,Oppedisano}.
Slow nucleons are classified from emulsion experiment into black
particles (the low energy target fragments emitted by evaporation) and
grey particles (the soft nucleons knocked out by wounded nucleons). The
features of emitted nucleons seem to depend weakly on the projectile
energy. This indicates that the emission of slow particles is mostly
dictated by nuclear geometry. Therefore we follow a parameterization
of results from low energy experiments, which describes the
proportionality of the soft nucleons to
$\ensuremath{N_\mathrm{coll}}$.

This model is coupled to the Glauber model which provides the
distribution of $\ensuremath{N_\mathrm{coll}}$. The results are shown
in Figure \ref{fig:glauFit}, on the right, where the Monte Carlo
Glauber + Slow Nucleon Model distribution are compared to the
distributions measured. The ZN-A spectrum is reasonably described by
the model.  Despite of saturation, which occurs at central events,
slicing the events with the ZN still provides a distinction in
centrality classes.

\begin{figure}[btp]
 \centering
  \includegraphics[width=0.75\textwidth]{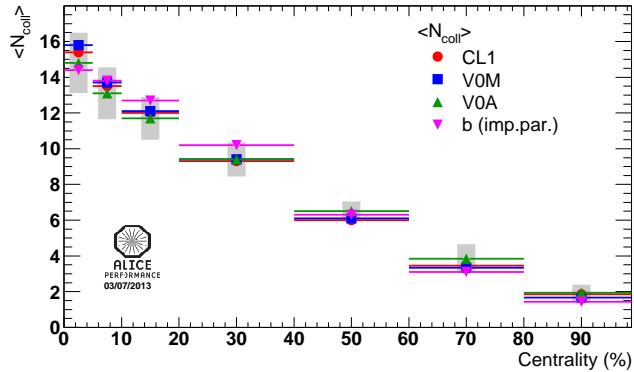}
  \caption{Values of $\ensuremath{N_\mathrm{coll}}$ extracted from CL1, V0M, V0A, ZNA and by slicing the hadronic cross-section from the impact parameter distribution.
  \label{fig:Ncoll}}
\end{figure}

\section{Bias in the centrality measurement}
In order to use these $\ensuremath{N_\mathrm{coll}}$ values in a $R_{pA}$
calculation, one needs to understand the bias arising when sampling
the pA events in centrality classes.  There is a much looser
correlation between impact parameter and
$\ensuremath{N_\mathrm{coll}}$, and between
$\ensuremath{N_\mathrm{coll}}$ (or $\ensuremath{N_\mathrm{part}}$) and
the charged particle multiplicity for p-Pb than for Pb-Pb
collisions. Also the correlation between different multiplicity
estimators is much broader in p-Pb than in Pb-Pb. The width of the
correlations demonstrates the importance of fluctuations, when
centrality is defined based on particle multiplicity, resulting in a
bias of the p-N collisions in a given centrality class.

\begin{figure}[btp]
 \centering
  \includegraphics[width=0.75\textwidth]{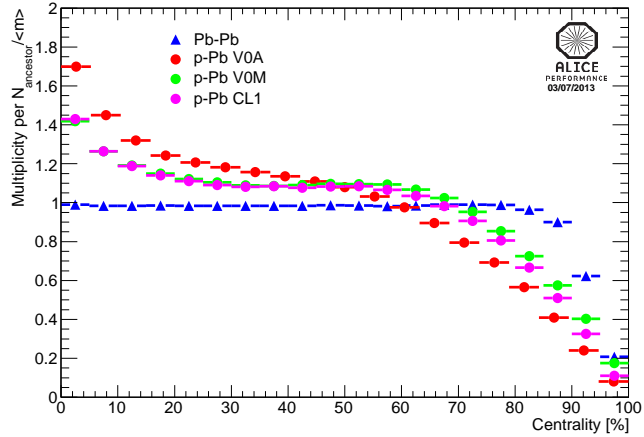}
  \caption{Multiplicity fluctuation bias calculated from the
    NBD-Glauber MC as the ratio between mean multiplicity per ancestor
    and the mean NBD multiplicity in p-Pb and Pb-Pb calculations.
  \label{fig:meanMult}}
\end{figure}

Directly from the NBD-Glauber fit an indication on the strength of
these bias can be derived.  Fig. \ref{fig:meanMult} shows the ratio of
the generated multiplicity per particle source divided by the mean
multiplicity of the NBD. In case of p-Pb, particle sources correspond
to $\ensuremath{N_\mathrm{part}}$, while for Pb-Pb, they are a
function of $\ensuremath{N_\mathrm{part}}$.  This ratio is constant in
case all collisions are unbiased. However the plot shows a function of
the centrality indicating a positive (negative) bias in central
(peripheral) events, much larger compared to what obtained in Pb-Pb
collisions, where the width of the plateau of the ancestor
distribution is large with respect to multiplicity fluctuations, and
only the most peripheral events are biased, due to their small
multiplicities.

To interpret this result Monte Carlo generators that correctly
simulate multi-particle production in nucleon-nucleon collisions are
needed. In all recent Monte Carlo generators, for example HIJING
\cite{Hijing}, a large part of the multiplicity fluctuations is due to
the fluctuations of the number of particle sources via multiple parton
interactions (MPI). Therefore, the biases on the multiplicity
corresponds to a bias on the number of hard scatterings in the
event. For peripheral (central) collisions we expect a lower (higher)
than average number of hard scatterings corresponding to a nuclear
modification factor $R_{pA} (\ensuremath{p_{\rm T}}) < 1$ ($R_{pA} (\ensuremath{p_{\rm T}}) > 1$).

In most cases, the MPI probability is governed by the impact parameter
between two nucleons. The mean nucleon-nucleon impact parameter as a
function of the number of participants can also be directly obtained
from a Monte-Carlo Glauber simulation and is shown here. For
collisions down to $\ensuremath{N_\mathrm{part}} \approx$4, it is constant
but increases significantly for peripheral collisions.  In peripheral
collisions, the multiplicity bias which gives a lower (higher) than
average number of hard scatterings for peripheral (central) collisions
is further enhanced by the higher than average nucleon-nucleon impact
parameter, that reduces the probability for MPI.

Concerning the nuclear modification factor $R_{pA}$ at high transverse
momentum another type of bias has to be discussed. This affects
only the most peripheral events. High momentum particles are
produced in the fragmentation of partons produced in parton-parton
scattering with large momentum transfer. These fragmentation products
contribute to the overall event multiplicity and this can introduce a
trivial correlation between the centrality estimator and the presence
of a high momentum particles in the event.  Specifically, a cut of the
most peripheral p-Pb collisions (e.g. 80-100\%) selects a range in
multiplicity smaller than the multiplicity range covered in pp,
therefore resulting in an effective veto on the large multiplicity
events produced by hard processes. Here, the multiplicity estimator
acts also as a veto on hard processes which contribute to the overall
multiplicity (jet-veto).

\section{Implications for binary scaling}
Practically for the different estimators used we expect different
deviations from $\ensuremath{N_\mathrm{coll}}$ scaling, namely:
\begin{itemize}
\item CL1 (Clusters Pixel Layer 2): strong bias due to full overlap
  with tracking region. Additional bias in peripheral event from Jet
  veto effect since jets contribute to the multiplicity and shift
  events to higher centralities ($\ensuremath{p_{\rm T}}$- dependent).
\item V0M (V0A+V0C) Multiplicity: reduced bias since it is outside the
  tracking region
\item V0A Multiplicity: reduced bias because of the important
  contribution from Pb fragmentation region.
\item ZNA: small bias since the slow nucleon production is
  independent of hard processes.
\end{itemize}

In general, the number of binary collisions
$\ensuremath{N_\mathrm{coll}}$ is used to scale the reference pp
yields and obtain the nuclear modification factor. However, from the
discussion above, it is expected that observables measured in
centrality classes based on particle multiplicity deviate from binary
scaling leading, at high momenta, to nuclear modification factors
$R_{pPb} < 1$ ($R_{pPb} > 1$) in peripheral (central)
collisions. These effects are enhanced in peripheral events by the
jet-veto and the rising mean impact parameter.  Therefore we define
$Q_{pA}$ for "centrality" selected data, as biased ratios, therefore
to be denoted by Q, not R, and constructed as: $Q^{range;EST}_{pA} =
\ensuremath{Yield(pA)^{range;EST}} /<\ensuremath{N_\mathrm{coll}}
(Glauber;EST)> \times \ensuremath{Yield(pp)}$, each biased by the use
of the particular estimator EST for the event ordering.

\begin{figure}
 \centering
  \includegraphics[width=0.45\textwidth]{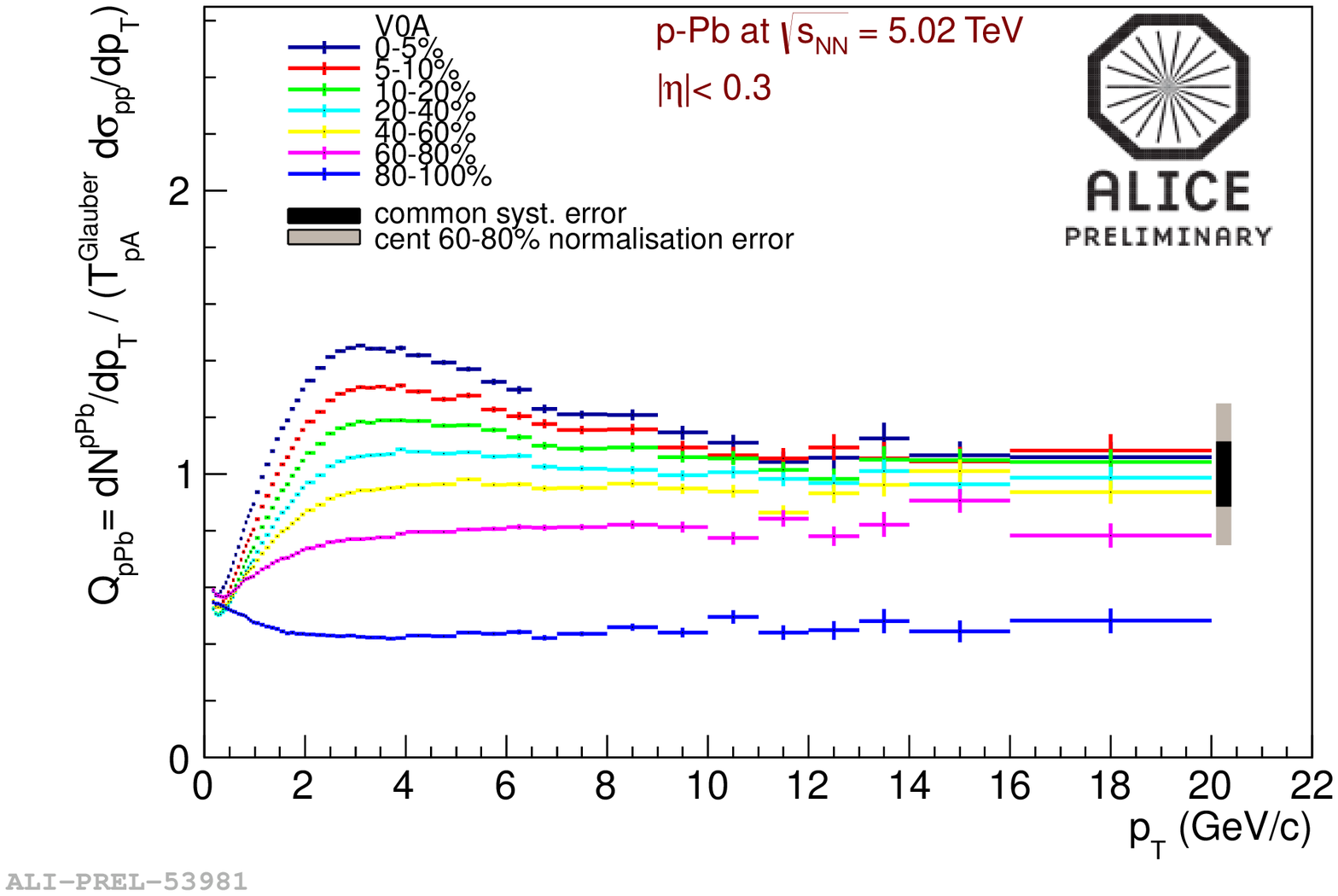}
  \includegraphics[width=0.45\textwidth]{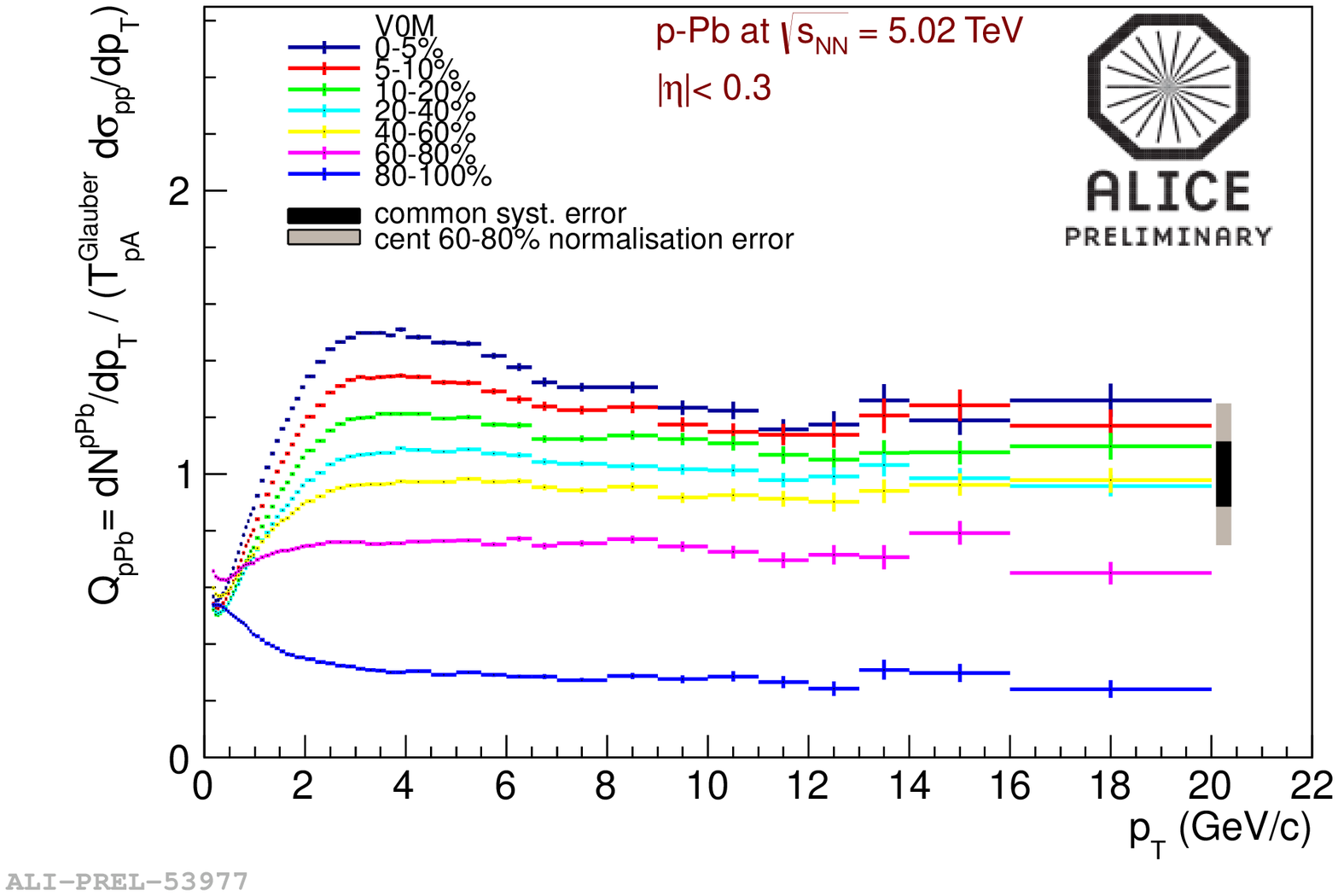}
  \includegraphics[width=0.45\textwidth]{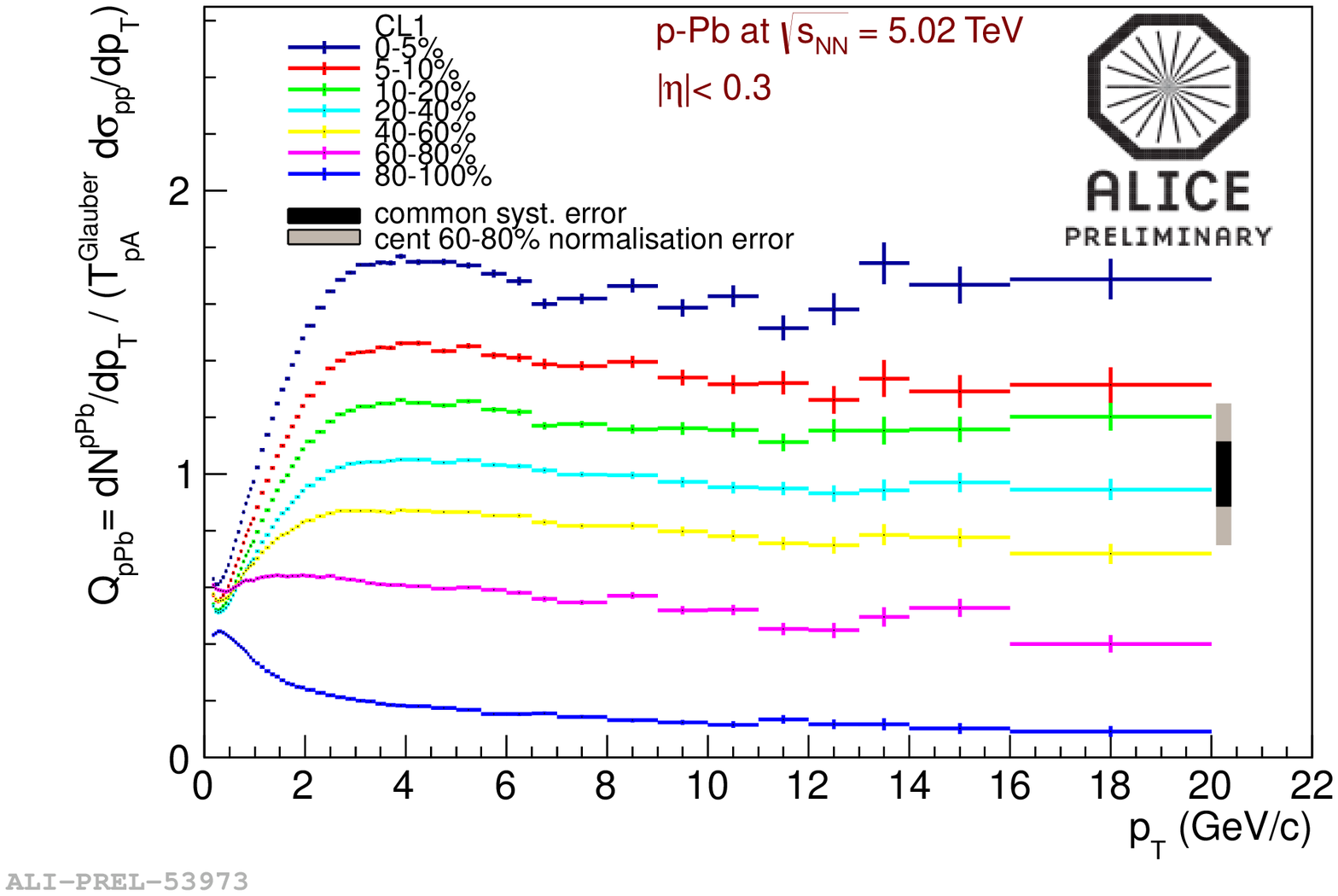}
  \includegraphics[width=0.45\textwidth]{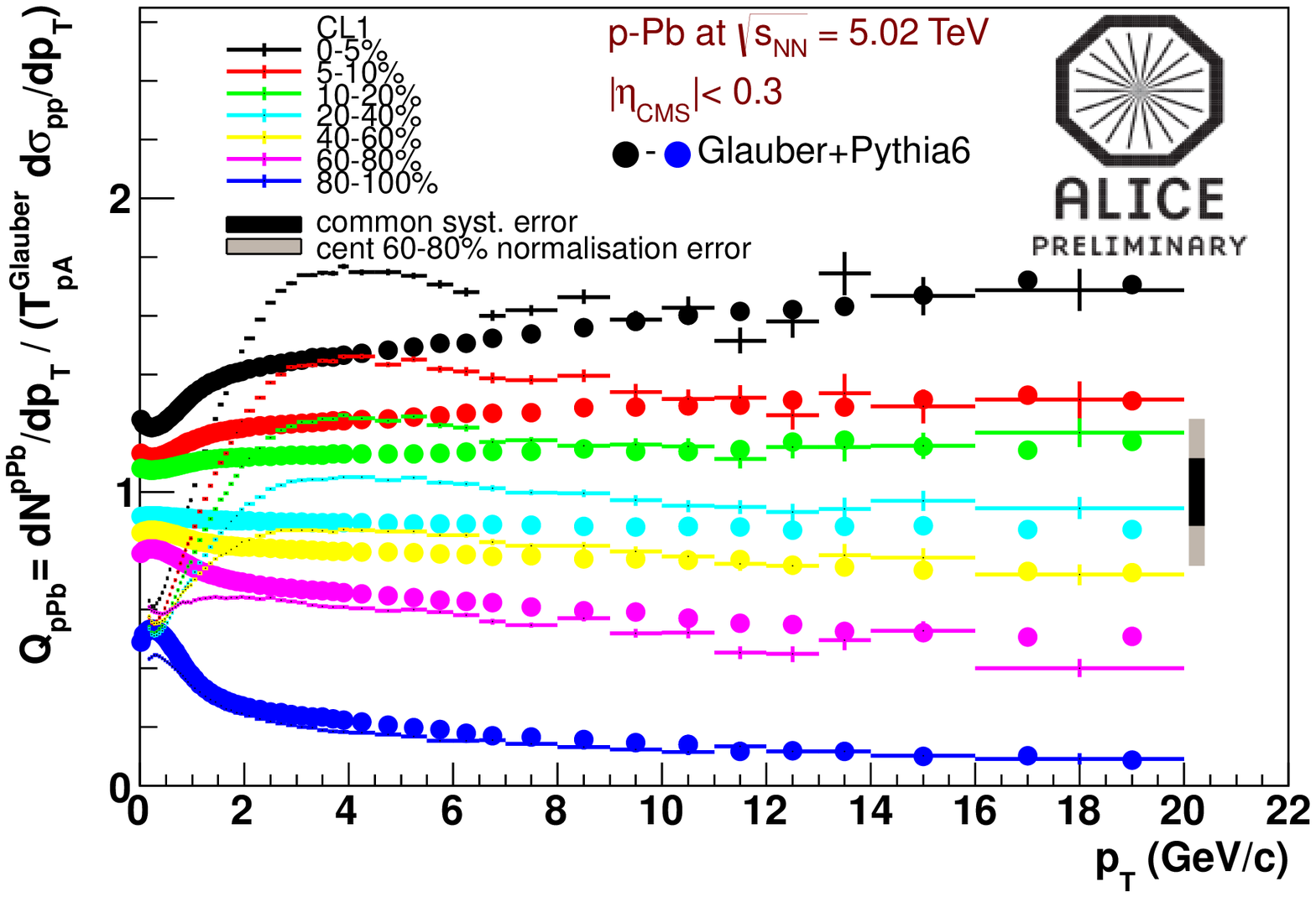}
  \caption{$Q_{pA}$  of all charged particles for various centrality classes obtained with different centrality estimators: V0A (top left), V0M (top right), CL1 (bottom left). Bottom right: The spectra for CL1 are compared to the one obtained with the Toy-MC (Pythia + Glauber).
  \label{fig:Qpa}}
\end{figure}

Figure \ref{fig:Qpa} shows the $Q_{pA}$ for different centrality
estimators and different centrality classes. For all centrality
classes $Q_{pA}$ strongly deviates from unity at high
$\ensuremath{p_{\rm T}}$, with values at high $\ensuremath{p_{\rm T}}$
well above unity for central collisions and below unity for peripheral
collisions. However the spread between centrality classes reduces with
increasing rapidity gap between the centrality estimator and the
$\ensuremath{p_{\rm T}}$ measurement, namely it is largest for CL1,
reduces for V0M and even further for V0A. The smallest bias is present
for the ZNA measurement. There is a clear indication for the jet-veto
bias in the most peripheral CL1 class $Q_{pA}$ has a significant
negative slope due to the fact that the contribution of jets to the
overall multiplicity increases with $\ensuremath{p_{\rm T}}$. This
jet-veto bias diminishes in V0M and is absent in V0A, where however
$Q_{pA}<1$, indicating that the multiplicity bias is still
present. The smallest bias is expected from the ZNA, the analysis of which is still in progress.

The mean $Q_{pA}$ at high $\ensuremath{p_{\rm T}}$ with the various
estimators shows the same centrality dependence as seen in
multiplicity bias (NBD-Glauber fit). If we model p-Pb collisions as
incoherent superposition of nucleon-nucleon by coupling in a Toy-MC a
Pythia \cite{pythia} simulation to our Glauber calculation, we can estimate the
trend of hard scatterings per collision as a function of
centrality. This multiplicity bias shows a strong deviation from
$\ensuremath{N_\mathrm{coll}}$-scaling at low and high centralities
with dependence very similar to the mean $Q_{pA}$ at high momenta. The
shape flattens with increasing rapidity gap, going from CL1 to V0M and
finally to V0A.

Therefore we compare the measured $Q_{pA}$ for the CL1 estimator to
the one obtained with the Toy-MC (Pythia + Glauber). The bias at high
$\ensuremath{p_{\rm T}}$ is described by incoherent superposition of
pp collisions, for all centralities. For most peripheral, there is
good agreement also in low- and intermediate $\ensuremath{p_{\rm T}}$
region. But at low- and intermediate $\ensuremath{p_{\rm T}}$ region
strong deviations are observed for all other centrality bins, which 
can be presumably attributed to nuclear modification effects, observed
in other physics observables.

\section{Conclusions}
In summary, centrality estimators based on multiplicity measurements
in $|\eta| < 5$ induce a bias on the hardness of the pN collisions
that can be quantified by the number of hard scatterings per pN
collision. Low (high) multiplicity p-Pb leads to lower (higher) than
average number of hard scatterings.  This bias can be quantified with
a toy MC that builds an incoherent superposition of pN collisions and
nuclear modification effects should be searched including this
bias. For "centrality" selected data, for which
$<\ensuremath{N_\mathrm{coll}}>$ is not uniquely defined, we
introduced $Q_{pA}$'s, each biased by the use of the particular
estimator for the event ordering. A selection based on the ZDC
provides a measurement with minimal (or absent?) bias, for which
should be possible to calculate an unbiased $R_{pA}$. There are
on-going efforts in the ALICE Collaboration to understand
$\ensuremath{N_\mathrm{coll}}$ from Slow Nucleon Model.

\end{document}